# Memristor Applications: Nanodevices Redefining Technological Landscapes

Nikolaos Vasileiadis
Institute of Nanoscience and Nanotechnology, NCSR "Demokritos
Ag. Paraskevi 15341, Greece
n.vasiliadis@inn.demokritos.gr

Georgios Ch. Sirakoulis
Department of Electrical and Computer Engineering, Democritus University of Thrace, Xanthi 67100, Greece
gsirak@ee.duth.gr

Panagiotis Dimitrakis
Institute of Nanoscience and Nanotechnology, NCSR "Demokritos",
Ag. Paraskevi 15341, Greece
p.dimitrakis@inn.demokritos.gr

*Abstract*— A memristor, a two-terminal nanodevice, has garnered substantial attention in recent years due to its distinctive properties and versatile applications. These nanoscale components, characterized by their simplicity of manufacture, scalability in small dimensions, nonvolatile memory capabilities, and adaptability to low-power platforms, offer a wealth of opportunities for technological innovation. Memristors hold great promise in diverse fields, ranging from advanced memory devices and neuromorphic computing to energy-efficient circuits and more. As we delve into this report, our aim is to provide a succinct but thorough exploration of the expanding landscape of memristor applications. Through the meticulous examination of scholarly literature, we systematically documented pivotal research milestones. By preserving historical consistency in our approach, we aim to unveil the intricate spectrum of possibilities that memristors offer, according to which they can revolutionize and enhance various domains of electronics and computing.

## I. INTRODUCTION

The challenge of memory storage in electronic systems has persisted as a fundamental problem over several decades. Memristors are promising candidates for replacing current nonvolatile memories and realize storage class memories. It was initially sparked by Chua's theoretical proposal of the memristor as a fourth essential passive circuit element [1]. Recent advancements in oxide-based resistance memory and memristor technologies have shown promise in surpassing the limitations associated with Si-based flash memory [2]. A significant breakthrough has emerged in the form of a TaOx-based asymmetric passive switching device. This innovation boasts remarkable attributes such as high-density storage, rapid switching, endurance exceeding $10^{12}$ cycles, and minimal power consumption, marking a substantial leap forward in addressing the memory storage dilemma [2]. In parallel, research into silicon nitride-based resistive switching memories (see Fig. 1) has shed light on the role of defects and oxygen doping. These insights offer a deeper understanding of the operational mechanisms of memristors, making them potential candidates for emerging non-volatile memory (NVM) technologies [3]. Table 1 summarizes the primary benefits and challenges associated with the four significant emerging NVMs [4]. Among these emerging NVMs, the FeFET stands out as an upcoming solution with a single transistor, offering a promising low-power NVM alternative. PCM, on the other hand, is the most advanced developing NVM with demonstrated performance and potential for further improvement in switching power reduction. To facilitate its early commercialization, significant research and development efforts have concentrated on enhancing materials and processes, given that STTRAM achieves maximum performance. Finally, RRAM holds the advantage of simplicity and potential cost-effectiveness. However, its reliability still requires substantial improvement.

Subsequently, as a result of extensive and expeditious advancements in memristor technology research and its multifarious applications, it has been demonstrated that the memristor not only serves the primary purpose of mitigating storage-related challenges but also offers remedies for a diverse array of other applications. In this work, an effort was undertaken to categorize the application spectrum of memristors and subsequently methodically record significant research milestones within each of these domains. Each of these domains is scrutinized in distinct sections of this study.

| Table 1 | Summary of advantages and challenges of major emerging NVMs. ||
|---|---|---|
| | *Main advantages* | *Key challenges* |
| FeFET | • IT cell structure<br>• Low-power field-driven<br>• High performance<br>• Ferroelectric doped HfO | • Material and processing<br>• FEOL integration<br>• Reliability and parasitic effects (e.g. charge trapping) |
| PCM | • Maturity<br>• Proven performance | • Reliability<br>• Disturbance<br>• High switching power |
| STTRAM | • High performance<br>• Well-understood physics<br>• Novel mechanisms (e.g., SHE, VCMA) to extend capabilities | • Reducing I/A (power-stability tradeoff)<br>• MTJ patterning and etching<br>• BEOL thermal budget |
| RRAM | • Simplicity and low cost<br>• High density<br>• Versatile materials, structures, and behaviors | • Reliability and failures<br>• Stochastic mechanism and intrinsic variability<br>• Forming |

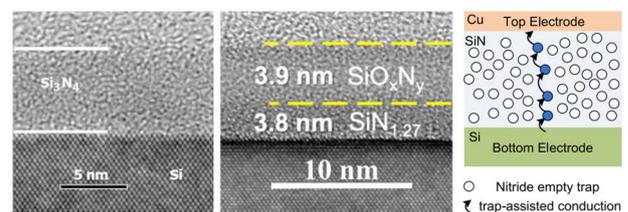

Fig. 1. The SiNx Memristor along with its mechanism, adopted from [3]

## II. THE VON-NEUMANN PROBLEM

The von Neumann bottleneck is a persistent challenge in computer architecture, arising from the disparity between CPU speed and slower memory access. It stems from the von Neumann architecture, where programming instructions and data share memory, unlike the separate Harvard architecture.

As computing evolved, efficient memory access became crucial, leading to various proposed solutions. These include caching for faster memory access, multithreading for process management, and modifications to memory bus design and parallel processing to boost memory bandwidth. Radical concepts like "non-von Neumann" systems inspired by biology aim to distribute memory intake differently. Emerging technologies like memristors are gaining attention for their potential to transcend traditional computing limits and address the von Neumann bottleneck. Memristors combine memory and resistance properties, offering innovative computing solutions and new approaches to tackling this challenge. In [5], researchers demonstrate the memristor's capability to perform material implication (IMP), a fundamental Boolean logic operation. This showcases how memristors can serve as both gates for logic operations and latches for memory, using resistance as the physical state variable. This insight paves the way for "stateful" logic operations, where memristors play a dual role in computing. Building on this foundation, [6] introduces a novel logic circuit design paradigm that leverages the threshold-dependent switching response of single memristors. This approach enables the creation of robust programmable composite memristive switches with variable precision. These switches are then applied in the design of memristive computing circuits, offering a new path to address the von Neumann bottleneck. Furthermore, [7] demonstrates the potential of memristors in pattern classification using artificial neural networks. By implementing a single-layer perceptron network with memristive crossbar circuits, researchers achieve efficient training and classification, despite variations in memristor behavior. This work holds promise for the realization of efficient neuromorphic networks and high-performance information processing systems. In [8], a fully integrated memristor-CMOS system is presented, facilitating efficient multiply-accumulate operations. This hybrid chip combines memristor crossbar arrays with custom-designed circuits, enabling online learning, vector-matrix operations, and mapping of various algorithms. This integration supports operational neuromorphic computing hardware, demonstrating its potential to address the von Neumann bottleneck and also paved the way for another crucial application of memristor technology which was found in [9], where researchers fabricate high-yield memristor crossbar arrays for implementing convolutional neural networks (CNNs). This advancement in hardware-based neural network training showcases significant energy efficiency gains compared to traditional graphics-processing units (GPUs), promising scalable solutions for deep neural networks and edge computing. Lastly, [10] introduces MemCA, an all-memristor design for deterministic and probabilistic cellular automata hardware. By incorporating memristors into both the cell and rule modules, this innovative approach enhances computational capabilities. MemCA demonstrates high speed, reduced area requirements, and adaptability, making it a promising candidate to tackle the von Neumann bottleneck through novel computing paradigms.

III. THE RECONFIGURABILITY PROBLEM

The field of reconfigurable electronics has experienced remarkable progress, marking a transformative phase in electronic engineering. Reconfigurable electronics encompasses the development of electronic systems and devices that possess the remarkable capability to dynamically adapt and reconfigure their hardware configurations to perform a diverse range of functions or tasks. This field's evolution has been underscored by pioneering developments, and at its core, memristor technology stands as a game-changing innovation. Beginning with the initial contribution [11] in the field, the authors harnessed the memristive properties of vanadium dioxide to create an adaptive filter within an LC circuit. This innovation enabled the circuit to respond dynamically to input signals, enhancing its resonant response quality factor. Additionally, mathematical extensions introduced memory-reactive elements like memcapacitors and meminductors, broadening the adaptive memory filter's functionality. Building upon [11], [12] introduced a memristor crossbar architecture for adaptive synchronization. Overcoming the challenge of massive interconnections, this paper demonstrated the effectiveness of this architecture, showcasing its robustness to device variability and faults. It harnessed memristors to synchronize nonlinear circuits, mirroring mechanisms observed in biological systems. Advancing to [13], memristive field-programmable analog arrays (memFPAA) were introduced as versatile analog computing platforms with reconfigurability. Memristive devices served as core analog elements, allowing the memFPAA to function as a bandpass filter, audio equalizer, and acoustic mixed frequency classifier. This versatile platform opens doors for rapid prototyping and efficient analog applications. Finally, [14] adheres to current state-of-the-art. A self-adaptive hardware system was proposed, incorporating resistive-switching synaptic arrays of memory devices. This system, driven by homeostatic Hebbian learning, showcased potential applications in reinforcement learning tasks, such as Mars rover navigation illuminating the evolution of memristor-based solutions in the context of reconfigurability within electronics.

IV. THE QUANTUM PROBLEM

The advent of quantum computing represents a pivotal milestone in the ongoing second quantum revolution, promising unparalleled computational capabilities. Quantum algorithms have exhibited remarkable prowess in tackling intricate problems, such as prime number factorization and unstructured database searches, with exceptional speed and efficiency. However, the formidable hurdle impeding the realization of large-scale, efficient quantum computers is the specter of decoherence. Decoherence introduces errors that necessitate continuous correction through the implementation of quantum error correcting codes. Additionally, quantum algorithms primarily rely on quantum logic gates, translatable and employable by classical computers, with interfaces grounded in linear algebra operations. Addressing this predicament, we turn to memristor technology, as proposed in [15, 16], which presents a groundbreaking solution. Memristive grids, as novel nanoscale, low-power hardware accelerators, emerge as a compelling avenue for expediting the computationally intensive matrix-vector multiplications and tensor products inherent in quantum computations (see Fig. 2). Leveraging the distinctive attributes of memristive grids, this work pioneers their deployment in circuit-level quantum computations. By recognizing that all quantum computations can be mapped to quantum circuits, memristive grids not only serve as efficient quantum simulators but also act as crucial components in classical-quantum computing systems, serving as interfaces and accelerators alike.

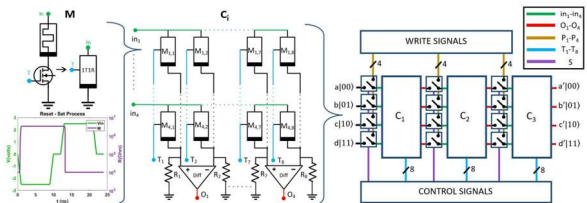

Fig. 2. Memristor based crossbar configuration that can compute quantum algorithms, adopted from [15]

## V. THE ENERGY AND PORTABILITY PROBLEM

The burgeoning demand for low-power and portable electronic devices in various applications has spurred significant interest in harnessing memristor technology to address these challenges. Memristive devices, known for their capacity to retain resistance states based on the history of applied voltage and current, hold the potential to revolutionize this aspect of electronics [17]. Analog computing utilizing memristors presents an energy-efficient alternative to conventional digital computing, particularly in the context of the Internet of Things (IoT) [18]. The IoT landscape necessitates processing vast amounts of real-time data at the edge, where conventional digital technologies struggle due to their need to convert analog signals into digital data. Memristors can streamline this process by enabling efficient processing of analog signals directly [18]. Furthermore, the integration of memristors into computing-in-memory (CIM) structures offers substantial improvements in energy efficiency and low-latency operation [19]. A fully integrated memristive CIM structure, demonstrated on a 65nm CMOS process, exhibits remarkable energy efficiency and performance for Boolean logic and multiply-and-accumulate operations [19]. This advancement could revolutionize AI edge processors by significantly reducing energy consumption while maintaining high computing speeds. In the realm of sensory networks, the concept of near-sensor and in-sensor computing emerges as a pivotal solution to reduce data movement and power consumption [20]. By offloading computation tasks closer to the sensory terminals, redundancy in data exchange can be minimized. This approach is poised to optimize the performance of sensory networks by utilizing memristor-based computing units and advanced manufacturing technologies [20]. Silicon nitride memristor technology aligns with the transformative paradigm of edge computing but also enhances it by seamlessly integrating memristors with photodiodes [21] or phototransistors [22]. This integration (see Fig. 3) offers a substantial advantage by expanding dynamic range of the sensor, courtesy of the larger resistive window inherent to SiNx memristors. This expanded resistive window plays a pivotal role in bolstering the performance of such sensor technology, thus underscoring the critical significance of SiNx memristors in this context. Additionally, in-sensor reservoir computing systems, comprising memristor arrays and deep ultraviolet (DUV) photo-synapses, hold promise for efficient latent fingerprint recognition, circumventing the need for separate memory and processors [23]. This innovation significantly reduces latency and overall computing power requirements. Lastly, wearable in-sensor reservoir computing systems leveraging optoelectronic polymers have the potential to enable multi-task learning, mimicking biological vision while minimizing time and energy overheads [24]. These systems, which incorporate memristive organic diodes in their readout functions, excel in recognizing various tasks with remarkable accuracy, providing a cost-effective and efficient solution for wearable devices.

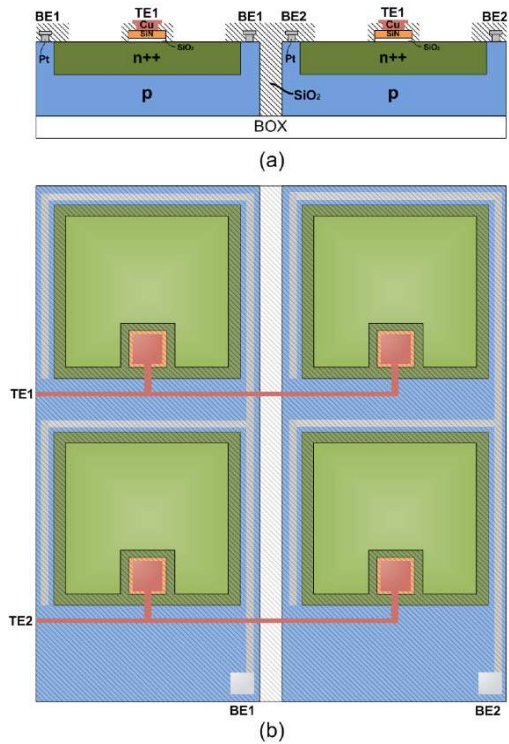

Fig. 3. 1D1M vision sensor's schematic layout, adopted from [21]

## VI. THE SECURITY PROBLEM

The proliferation of electronics and the increasing need for secure information processing have spurred research in the realm of true random number generators (TRNGs). Traditional cryptographic methods, while effective, face challenges related to device cloning and secret key extraction. Recent advancements in nanoelectronic devices, particularly memristors (as seen in [25] and [26]), have offered promising solutions. Memristive devices exploit process variations to develop physically unclonable functions (PUFs), providing uniqueness, reliability, and a large number of challenge-response pairs. Notably, some memristor-based PUFs (mrSPUFs) can serve as reconfigurable PUFs (rPUFs) without additional hardware or aiding applications that requiring secure key updates. Additionally, the utilization of conventional CMOS technology in [27] presents a True Random Number Generator that harnesses the randomness of telegraph noise in a single CMOS transistor, achieving enhanced complexity, output bit rate, and power efficiency. Robustness against machine learning attacks further solidifies its security. In [28], silicon nitride memristors are employed to create a TRNG, benefiting from multi-state currents as entropy sources, while a simple Xorshift logic circuit post-processes the bitstreams (see fig. 4). This design aligns with the demands of smaller, faster, and more power-efficient TRNGs crucial in the era of the Internet of Things. Lastly, [29] introduces a novel approach, leveraging post-CMOS memristor devices as a fingerprinting method to uniquely identify integrated circuits (ICs). By exploiting memristors' variable I-V characteristics, this methodology offers a generally applicable, counterfeit-resistant solution for supply chain tracking and quality assurance. Together, these contributions highlight the evolving landscape of TRNGs and their potential to address critical security challenges in modern electronics.

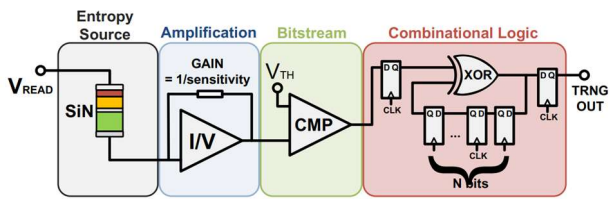

Fig. 4. The Memristive Xorshift True Random Number Generator Circuit, adopted from [28].

## VII. THE ELECTROMECHANICAL PROBLEM

High-performance switches play a crucial part in the field of radiofrequency (RF) circuits, critical for applications in wireless communications and consumer electronics. As conventional CMOS technologies face inherent limitations related to speed, energy consumption, and temperature constraints. Additionally, their physical size impedes seamless integration with nanometer-scale CMOS RF circuits. A promising solution [30] introduced a nanoscale RF switch based on a nonvolatile memristive device. This innovative design features electrochemically asymmetric metal electrodes with a nanometer-scale air gap, yielding low ON-state resistance and low OFF-state capacitance, crucial for achieving high cutoff frequencies and maintaining linearity in wide-band systems. Leveraging silver (Ag) as the active metal and silicon oxide for the air gap results in a low effective dielectric constant. This approach delivers remarkable performance metrics, including a low insertion loss of 0.3 dB at 40 GHz, high isolation of 30 dB at 40 GHz, and a typical cutoff frequency of 35 THz, all while preserving competitive linearity and power-handling capabilities. Also, the latest contribution in the field explored the design of a differential Single-Pole Double-Throw (SPDT) switch (see fig. 5), emphasizing its suitability in wireless transceivers and RF/microwave applications. The assessment includes critical figures-of-merit and strategies for mitigating added noise, with particular attention to addressing limitations associated with output subtractors [31].

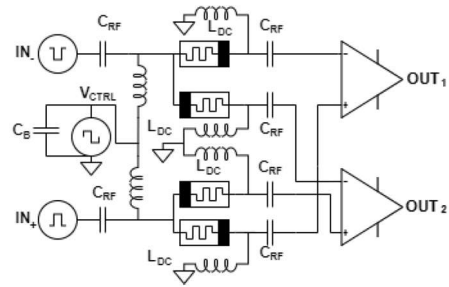

Fig. 5. Schematic depiction of the proposed DPDT memristive switch utilized as a SPDT switch, adopted from [31].

## VIII. THE SYNAPSE EMULATION PROBLEM

The issue of plasticity in electronics, particularly in the context of biological and neuromorphic systems, has been a subject of extensive research over the years, spanning from the early exploration of memristive models to the most recent advancements in the field. Initially, the Memristive model of amoeba's learning [32] sought to unravel the mechanisms underlying the remarkable behavioral intelligence observed in the Physarum polycephalum. This study established a connection between the amoeba's learning behavior and a simple electronic circuit featuring a memristor, shedding light on the microscopic origins of memristive behavior within biological organisms. As interdisciplinary research evolved, the intersection of technology and neuroscience led to the recognition of the memristor's potential in understanding synaptic plasticity. Memristance can explain Spike-Time-Dependent-Plasticity in Neural Synapses [33] emerged as a groundbreaking exploration, demonstrating how memristance models can naturally manifest Spike-Time-Dependent-Plasticity when combined with neural impulse signals. This discovery offers insights into the molecular and physiological mechanisms behind synaptic plasticity and paves the way for self-adaptive, intelligent machines. Fast-forwarding to contemporary developments, a neuromorphic physiological signal processing system based on memristor for next-generation human-machine interface [34] leverages memristors with their volatile threshold switching characteristics to construct an efficient neuromorphic system for processing physiological signals. This innovation showcases the potential of memristors in enhancing human-machine interfaces, achieving remarkable accuracies in arrhythmia classification and epileptic seizure detection. Furthermore, addressing the complexities of neurological systems, Hardware Design of Memristor-based Oscillators for Emulation of Neurological Diseases [35] introduces Memristor-based Oscillators (MBOs), which seamlessly integrate artificial neurons and synapses. This transistor-free and highly integrated circuit design utilizes passive unipolar memristor devices, showcasing their ability to replicate bio-plausible spiking and bursting activities. These MBO neurons hold promise for low-cost emulation of neurological diseases, contributing to our understanding of complex neural networks.

ACKNOWLEDGMENT

This work was supported in part by the research projects "3D-TOPOS" (MIS 5131411) and "LIMA-chip" (Proj.No. 2748) which are funded by the Operational Programme NSRF 2014-2020 and the Hellenic Foundation of Research and Innovation (HFRI) respectively.